# Variance Estimation for the Inverse Probability of Treatment Weighted Kaplan–Meier Estimator

Zhiwei Zhang[1], Yongwu Shao[1,2], Zhishen Ye[1]

## Introduction

In a widely cited paper, Xie and Liu [1] (henceforth XL) proposed to use inverse probability of treatment weighting (IPTW) to account for possible confounding in observational studies with survival endpoints subject to right censoring. Their proposal includes an IPTW Kaplan-Meier (KM) estimator for the survival function of a treatment-specific potential failure time, which can be used to evaluate the causal effect of one treatment versus another. The IPTW KM estimator is remarkably simple and highly effective for confounding bias correction, as demonstrated by XL and many others.  The method has been implemented in SAS's popular procedure LIFETEST for analyzing survival data and has seen widespread use.  We commend XL for their innovative and impactful work.

This letter is concerned with variance estimation for the IPTW KM estimator. The variance estimator provided by XL does not account for the variability of the IPTW weight when the propensity score is estimated from data, as is usually the case in observational studies. Indeed, their variance estimator in Proposition 2 remains the same whether the true propensity score is known or not. It is well known that the asymptotic variance of an inverse probability weighted estimator generally decreases as a result of estimating the weight [2]. In the specific context of randomized trials, IPTW has been shown to improve efficiency for treatment effect estimation when the propensity score, a known constant, is estimated by fitting a logistic regression model [3, 4]. Thus, the IPTW KM estimator is expected to become more efficient when the weight is based on an estimated propensity score as opposed to the true value. The XL variance estimator does not acknowledge this efficiency benefit of estimating the propensity score, and the resulting inference is expected to be conservative in these scenarios.

In this letter, we provide a rigorous asymptotic analysis of the IPTW KM estimator based on an estimated propensity score. Our analysis indicates that estimating the propensity score does tend to result in a smaller asymptotic variance, which can be estimated consistently using a plug-in variance estimator. We also present a simulation study comparing the variance estimator we propose with the XL variance estimator. Our simulation results confirm that the proposed variance estimator is more accurate than the XL variance estimator, which tends to over-estimate the sampling variance of the IPTW KM estimator.

---

[1] Biometrics, Gilead Sciences, Foster City, CA, USA
[2] Correspondence, E-mail: ywshao@gmail.com

## Variance Formula for the IPTW KM Estimator

We consider an observational study with a survival endpoint subject to right censoring. For a generic subject in the study, let $T$ be the observed (failure or censoring) time, $\delta$ the censoring indicator (1 observed; 0 censored), $X$ a treatment indicator (1 treated; 0 control), and $Z$ a vector of baseline covariates whose first component is a constant variable of 1. We assume that IPTW weights are obtained by fitting a logistic regression model:

$$\Pr(X = 1|Z) = \exp(\gamma_0^T Z)/[1 + \exp(\gamma_0^T Z)],$$

where $\gamma_0$ is a vector of true coefficient values. The objective is to estimate $S^{(k)}(t)$, $k = 0,1$, the potential survival function that would result from applying treatment $k$ to all individuals in the study population. Additional notations are defined in Table 1.

*Table 1. Notations*

| Notation | Definition |
|---|---|
| $O$ | $(T, \delta, X, Z)$ for a generic subject |
| $O_i$ | $(T_i, \delta_i, X_i, Z_i)$ for the $i$-th subject in the study |
| $\hat{\gamma}$ | Maximum likelihood estimate of $\gamma_0$ |
| $g(x)$ | $\exp(x)/[1 + \exp(x)]$ |
| $w(\gamma, O)$ | $1\{X = 1\}/g(\gamma^T Z) + 1\{X \neq 1\}/[1 - g(\gamma^T Z)]$ |
| $N_{ik}(t)$ | $I\{T_i \leq t, \delta_i = 1, X_i = k\}$ |
| $\hat{F}_k(t)$ | $(1/n)\sum w(\hat{\gamma}, O_i) N_{ik}(t)$ |
| $F_k(t)$ | $E[w(\gamma_0, O) N_{1k}(t)]$ |
| $Q_{ik}(t)$ | $I\{T_i \geq t, X_i = k\}$ |
| $\hat{Q}_k(t)$ | $(1/n)\sum w(\hat{\gamma}, O_i) Q_{ik}(t)$ |
| $Q_k(t)$ | $E[w(\gamma_0, O) Q_{1k}(t)]$ |
| $V_1$ | $E\{g(\gamma_0^T Z)[1 - g(\gamma_0^T Z)] ZZ^T\}$ |
| $\zeta(O)$ | $V_1^{-1} Z[X - g(\gamma_0^T Z)]$ |

Assume the survival time is continuous and there are no ties in the failure time. Let

$$\varphi_k(O_i) = S^{(k)}(t) \int_0^t \frac{Q_{ik}(v) dF_k(v) - Q_k(v) dN_{ik}(v)}{Q_k(v)^2}.$$

By Lemma 1 in the Appendix, we have

$$\hat{S}^{(k)}(t) - S^{(k)}(t) = \frac{1}{n}\sum_{i=1}^{n} w(\gamma_0, O_i)\varphi_k(O_i) + (\hat{\gamma} - \gamma_0)\mathrm{E}\left[\frac{\partial w(\gamma_0, O)}{\partial \gamma_0}\varphi_k(O)\right] + o_p(n^{-1/2}).$$

Part (b) of Proposition 2 in XL suggests that the standard error can be calculated by treating $\hat{\gamma}$ as fixed, which amounts to ignoring the second term in the preceding display. However, in general, $\mathrm{E}\left[\frac{\partial w(\gamma_0, O)}{\partial \gamma_0}\varphi_k(O)\right] \neq 0$, so the second term in the preceding display is of the order $O_p(n^{-1/2})$ and cannot be ignored as a lower order term. Since

$$\hat{\gamma} - \gamma_0 = \frac{1}{n}\sum_{i=1}^{n}\zeta(O_i) + o_p(n^{-1/2}),$$

where $\zeta(O)$ is defined in Table 1, we have

$$\hat{S}^{(k)}(t) - S^{(k)}(t) = \frac{1}{n}\sum_{i=1}^{n}\psi_k(O_i) + o_p(n^{-1/2}),$$

where

$$\psi_k(O) = w(\gamma_0, O)\varphi_k(O) + \zeta^T(O)\mathrm{E}\left[\frac{\partial w(\gamma_0, O)}{\partial \gamma_0}\varphi_k(O)\right].$$

Noting that the terms $\psi_k(O_j)$'s are independently identical distributed, the variance of the IPTW KM estimator is

$$\mathrm{Var}[\hat{S}^{(k)}(t)] = \mathrm{Var}[\psi_k(O)]/n + o(n^{-1}).$$

In practice, $\mathrm{Var}[\psi_k(O)]$ can be consistently estimated by the sample standard deviation of $\hat{\psi}_k(O_i)$ defined as below, which basically is a plug-in estimator of $\psi_k(O_i)$ with $\gamma_0$, $Q_k(v)$ and $F_k(v)$ replaced by their respective estimators $\hat{\gamma}$, $\hat{Q}_k(v)$, $\hat{F}_k(v)$.

$$\hat{\phi}_{1,k}(O_i) = \int_0^t \frac{Q_{ik}(v)d\hat{F}_k(v)}{\hat{Q}_k(v)^2} = \sum_{j:\delta_j=1}\frac{1\{T_j \leq \min(t, T_i), X_j = k\}w(\hat{\gamma}, O_j)}{n\hat{Q}_k(T_j)^2}$$

$$\hat{\phi}_{2,k}(O_i) = \int_0^t \frac{dN_{ik}(v)}{\hat{Q}_k(v)} = \frac{1\{T_i \leq t, \delta_i = 1, X_i = k\}}{\hat{Q}_k(T_i)}$$

$$\hat{\varphi}_k(O_i) = \hat{S}^k(t)[\hat{\phi}_{1,k}(O_i) - \hat{\phi}_{2,k}(O_i)]$$

$$\hat{V}_1 = \frac{1}{n}\sum_{i=1}^{n}g(\hat{\gamma}^T Z_i)[1 - g(\hat{\gamma}^T Z_i)]Z_i Z_i^T$$

$$\hat{\zeta}(O_i) = \frac{\hat{V}_1^{-1}}{n} \sum_{i=1}^{n} Z_i [X_i - g(\hat{\gamma}^T Z_i)]$$

$$\hat{\psi}_k(O_i) = w(\hat{\gamma}, O_i)\hat{\varphi}_k(O_i) + \frac{\hat{\zeta}^T(O_i)}{n} \sum_{i=1}^{n} \left.\frac{\partial w(\gamma, O_i)}{\partial \gamma}\right|_{\hat{\gamma}} \hat{\varphi}_k(O_i) \qquad (1)$$

The above formulas may look complicated; however, they can be implemented in R with <100 lines of code.

The causal effect of treatment 1 versus treatment 0 can be measured using the treatment difference $S^{(1)}(t) - S^{(0)}(t)$, which can be estimated as the difference between two IPTW KM estimates. For variance estimation and inference, we note that

$$\text{Var}\bigl[\hat{S}^{(1)}(t) - \hat{S}^{(0)}(t)\bigr] = \text{Var}[\psi_1(O) - \psi_0(O)]/n + o_p(n^{-1}).$$

Here, $\text{Var}[\psi_1(O) - \psi_0(O)]$ can be consistently estimated by the sample standard deviation of $\hat{\psi}_1(O_i) - \hat{\psi}_0(O_i)$.

## Simulations

We conducted a simple simulation study to evaluate the performance of our proposed standard error formulas given by (1) and by XL. We set $n = 1000$, $\tilde{Z} \sim \text{Normal}(0, I_3)$, $Z = (1, \tilde{Z})$, $\gamma_0 = (0, 0, 0.5, 1)^T$, and $X \sim \text{Bernoulli}(g(\gamma_0^T Z))$. The true survival time follows an exponential distribution with scale parameter $\exp(\eta_0^T Z)$, where $\eta_0 = (0, 1, 2, 3)^T \beta_0$ with $\beta_0$ ranging from 0 to 2. Larger values of $\beta_0$ represent stronger associations of $T$ with $Z$. The censoring variable follows an exponential distribution with scale parameter 1. We generate 1000 independent samples. For each sample, we calculate the IPTW KM estimate, as well as the standard error using (1) and XL's formula. We then calculate the Monte Carlo standard deviation of the estimates at $t = 0.5$ for the experimental group and the treatment difference, as well as the means of the standard errors from the two methods. The results are plotted in Figure 1 and Figure 2.

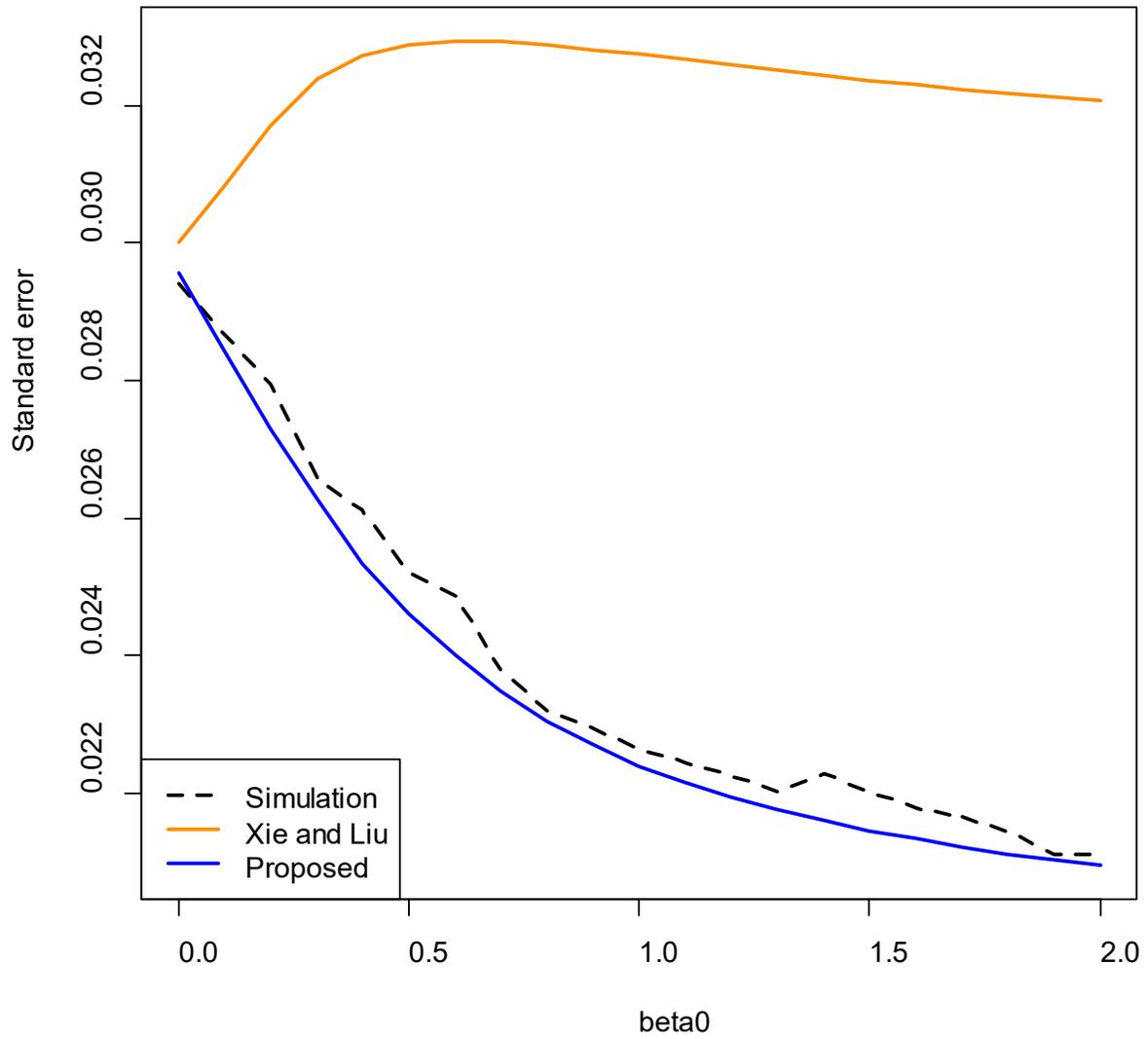

Figure 1. Estimated standard deviation of the IPTW KM estimate $\hat{S}^{(1)}(0.5)$ by simulation, and the standard errors calculated by the two methods (proposed, and the formula by XL).

*Figure 2. Estimated standard deviation of $\hat{S}^{(1)}(0.5) - \hat{S}^{(0)}(0.5)$ by simulation, and the standard errors calculated by the two methods (proposed, and the formula by XL).*

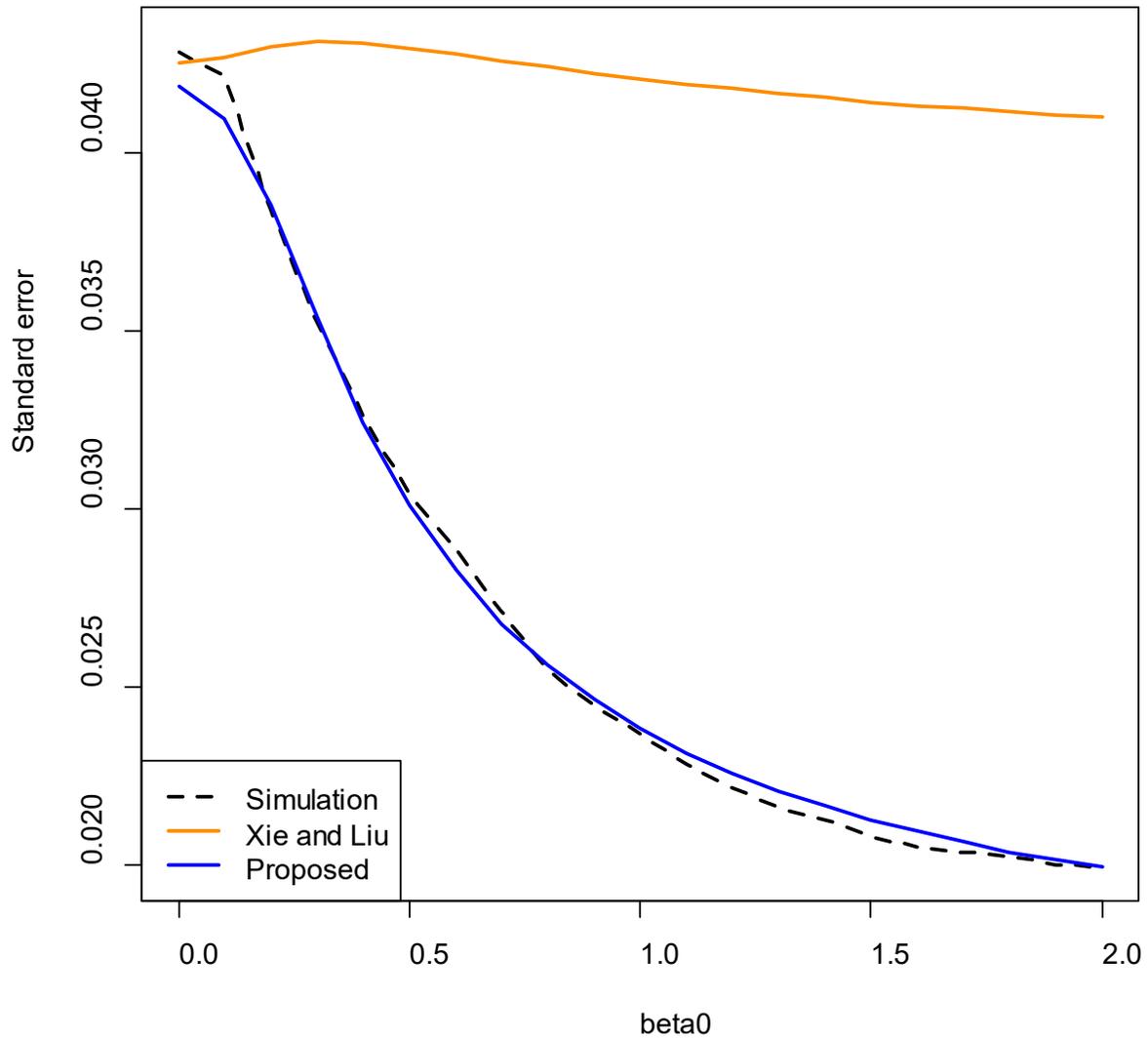

From the figures we can see that, as $\beta_0$ increases, the standard deviation of the IPTW KM estimates gradually decreases, which is expected as strong baseline predictors should help to reduce the variance. Our proposed variance estimator correctly predicts this trend, and it matches the actual standard deviation quite well, while XL's estimator tends to overestimate the standard deviation.

## Appendix

**Lemma 1.** $\hat{S}^{(k)}(t) - S^{(k)}(t) = \frac{1}{n}\sum_{i=1}^{n} w(\gamma_0, O_i)\varphi_k(O_i) + (\hat{\gamma} - \gamma_0)\mathrm{E}\left[\frac{\partial w(\gamma_0, O)}{\partial \gamma_0}\varphi_k(O)\right] + o_p(n^{-1/2})$.

*Proof.*

$$
\begin{aligned}
-\log \hat{S}^{(k)}(t) &= -\int_0^t \log\left[1 - \frac{1}{n\hat{Q}_k(v)}\right] d[n\hat{F}_k(v)] \\
&= \int_0^t \frac{d\hat{F}_k(v)}{\hat{Q}_k(v)} + o_p(n^{-1/2}) \\
&= \int_0^t \frac{d\hat{F}_k(v)}{Q_k(v)} - \int_0^t [\hat{Q}_k(v) - Q_k(v)]\frac{dF_k(v)}{Q_k(v)^2} + o_p(n^{-1/2}) \\
&= \int_0^t \frac{dF_k(v)}{Q_k(v)} + \frac{1}{n}\sum_{i=1}^n w(\hat{\gamma}, O_i) \int_0^t \frac{Q_k(v)dN_{ik}(v) - Q_{ik}(v)dF_k(v)}{Q_k(v)^2} + o_p(n^{-1/2}) \\
&= \int_0^t \frac{dF_k(v)}{Q_k(v)} + \frac{1}{n}\sum_{i=1}^n w(\hat{\gamma}, O_i)\,\varphi_k(O_i)/S^k(t) + o_p(n^{-1/2})
\end{aligned}
$$

Note the first term is simply $-\log S^{(k)}(t)$. Therefore, we have

$$
\begin{aligned}
\hat{S}^{(k)}(t) - S^{(k)}(t) &= \frac{1}{n}\sum_{i=1}^n w(\hat{\gamma}, O_i)\varphi_k(O_i) + o_p(n^{-1/2}) \\
&= \frac{1}{n}\sum_{i=1}^n w(\gamma_0, O_i)\varphi_k(O_i) + \frac{\hat{\gamma} - \gamma_0}{n}\sum_{i=1}^n \frac{\partial w(\gamma_0, O)}{\partial \gamma_0}\varphi_k(O_i) + o_p(n^{-1/2}) \\
&= \frac{1}{n}\sum_{i=1}^n w(\gamma_0, O_i)\varphi_k(O_i) + (\hat{\gamma} - \gamma_0)\mathrm{E}\left[\frac{\partial w(\gamma_0, O)}{\partial \gamma_0}\varphi_k(O)\right] + o_p(n^{-1/2})
\end{aligned}
$$

∎